\documentclass[12pt]{article}
\usepackage{epsfig,amssymb,amsmath,psfrag}

\catcode`\@=11
\newcommand{\insertfig}[2]{\mbox{\epsfxsize=#1cm \epsfbox{#2.eps}}}

\font\cmss=cmss12 
\def\1{\hbox{{1}\kern-.25em\hbox{l}}}
\def\bfZ{\relax{\hbox{\cmss Z\kern-.4em Z}}}

\def \be  {\begin{equation}}
\def \ee  {\end{equation}}
\def \ba  {\begin{eqnarray}}
\def \ea  {\end{eqnarray}}
\def \baa {\begin{eqnarray*}}
\def \eaa {\end{eqnarray*}}
\def \bb  {\begin {thebibliography} }
\def \eb  {\end{thebibliography}}
\def \lab #1 {\label{#1}}

\newcommand\re[1]{(\ref{#1})}

\def \matrix #1 {\left(\begin{array}{cc} #1 \end{array}\right)}

\def \tr {\mathop{\rm tr}\nolimits}

\newcommand{\as}{\ifmmode\alpha_{\rm s}\else{$\alpha_{\rm s}$}\fi}
\newcommand{\asbar}{\ifmmode\bar{\alpha}_{\rm s}\else{$\bar{\alpha}_{\rm s}$}\fi}

\newcommand{\bit}[1]{\mbox{\boldmath$#1$}}
\newcommand{\ft}[2]{{\textstyle\frac{#1}{#2}}}

\font\cmss=cmss12 
\def\inbar{\,\vrule height1.5ex width.4pt depth0pt}
\def\IC{\relax\hbox{$\inbar\kern-.3em{\rm C}$}}
\def\IZ{\relax{\hbox{\cmss Z\kern-.4em Z}}}
\def\IR{{\hbox{{\rm I}\kern-.2em\hbox{\rm R}}}}

\def\IP{{\hbox{{\rm I}\kern-.2em\hbox{\rm P}}}}
\def\II{\hbox{{1}\kern-.25em\hbox{l}}}

\def\numberbysection{\@addtoreset{equation}{section}
                     \def\theequation{\thesection.\arabic{equation}}}
\numberbysection

\newbox\lett\newdimen\lheight\newdimen\lwidth
\def\ontop#1#2{\setbox\lett=\hbox{#2}\lheight\ht\lett
\multiply\lheight by 12 \divide\lheight by 10\relax%
\lwidth\wd\lett \multiply\lwidth by 8 \divide\lwidth by 10\relax #2\kern-\lwidth%
\raise\lheight\hbox{{$\scriptstyle #1$}}\kern.1ex}

\def\XXint#1#2#3{{\setbox0=\hbox{$#1{#2#3}{\int}$}
     \vcenter{\hbox{$#2#3$}}\kern-.5\wd0}}

\begin{document}

\begin{titlepage}

\vskip2cm

\centerline{\large \bf Baxter equation for long-range $SL(2|1)$ magnet}

\vspace{1cm}

\centerline{\sc A.V. Belitsky}

\vspace{10mm}

\centerline{\it Department of Physics, Arizona State University}
\centerline{\it Tempe, AZ 85287-1504, USA}

\vspace{1cm}

\centerline{\bf Abstract}

\vspace{5mm}

We construct a long-range Baxter equation encoding anomalous dimensions of composite operators
in the $SL(2|1)$ sector of $\mathcal{N} = 4$ supersymmetric Yang-Mills theory. The formalism is
based on the analytical Bethe Ansatz. We compare predictions of the Baxter equations for short
operators with available multiloop perturbative calculations.

\end{titlepage}

\setcounter{footnote} 0

\thispagestyle{empty}

\newpage

\pagestyle{plain} \setcounter{page} 1

\section{Introduction}

With the discovery of integrable structures\footnote{Earlier, integrability was found to
emerge in reggeon interaction of high-energy scattering amplitudes \cite{Lip93,FadKor95}.}
in QCD \cite{BraDerMan98,BraDerKorMan99,Bel99} and maximally supersymmetric Yang-Mills theory
\cite{Lip98,MinZar03,BeiSta03,BeiKriSta03} it appears that understanding of strong coupling
behavior of anomalous dimensions of composite operators is within reach, at least in the latter
gauge theory. At one loop, the dilatation operators in the large$-N_c$ limit is identified with
a known Hamiltonian of a (graded) noncompact Heisenberg magnet with the quantum space in all
sites corresponding to infinite-dimensional representations of (super)conformal symmetry algebra
of gauge theory Lagrangians. While at higher loops it is mapped to yet to be determined putative
long-range spin chain. There are two generic approaches to integrable models, one based on (nested)
Bethe Ansatz \cite{Yan67} and another relying on the Baxter equation \cite{Bax80}. While both give
identical results for models based on representations with highest and/or lowest weight vectors,
the Baxter framework applies even when the pseudovacuum state in the Hilbert space of the chain is
absent. For noncompact super-spin chains, the number of eigenstates is infinite for a finite length
of the spin chain and the analysis of spectra in this approach is advantageous. The Bethe Ansatz
approach to multiloop dilatation operator in $\mathcal{N} = 4$ super-Yang-Mills theory was successfully
undertaken in Ref.\ \cite{BeiDipSta04,Sta04} culminating with conjectured long-range Bethe Ansatz
equations for the $PSU(2,2|4)$ spin chain \cite{BeiSta05}. The alternative formulation within the
framework of the Baxter equation was unavailable due to lack of efficient techniques to work out Baxter
equations for graded spin chains even with nearest-neighbor interactions. In Ref.\ \cite{BelDerKorMan06}
we have suggested a regular procedure to construction of the Baxter $\mathbb{Q}-$operators for
short-range $SL(2|1)$ magnet which is straightforwardly generalizable to supergroups of higher rank.
The $\mathbb{Q}-$operator which determines the energy spectrum of the chain, and thus the one-loop
anomalous dimensions of the dilatation operator in either holomorphic sector of $\mathcal{N} = 1$
super-Yang-Mills theory \cite{BelDerKorMan04} or in the minimally supersymmetric noncompact sector
of the maximally supersymmetric gauge theory \cite{Bei04}, was shown to obey a second order
finite-difference equation---the Baxter or TQ-equation---very similar to the one of the bosonic $SL(2)$
chain \cite{BraDerKorMan99,Bel99}. Recently we have suggested a long-range generalization of the Baxter
equation in the noncompact one-component $SL(2)$ sector of $\mathcal{N} = 4$ supersymmetric Yang-Mills
theory \cite{Bel06}. A next natural step is to extend the formalism to the noncompact graded $SL(2|1)$
subsector \cite{Bei04} of this gauge theory. Since the first principle microscopic formalism to build
long-range chains is currently unavailable, we reply on an effective approach using the analytical
Bethe Ansatz \cite{Res84}. The latter allows to determine the spectrum of transfer matrices from
known Bethe equations and as a result to obtain (nested) TQ-relations in terms of eigenvalues $Q$
of $\mathbb{Q}-$operators and eliminating the nested $Q-$functions determine the Baxter equations
\cite{BelKor05}.

\section{$SL(2|1)$ sector and quantum numbers of operators}

The $SL(2|1)$ symmetry arises as a reduction of the full superconformal symmetry group $SU(2, 2|4)$ of
the four-dimensional $\mathcal{N} = 4$ theory on the light-cone and operating on the complex scalar field
$X (z)$ and a single-flavor gaugino $\psi (z)$. Gauge theory leads to a particular realization of the
$SL(2|1)$ algebra on the space of functions of the light-cone chiral superspace $Z = (z, \theta)$.
Both fields can be accommodated into a single $\mathcal{N} = 1$ chiral superfield
\be
\Phi (Z) = i X (z) + \theta \psi (z)
\, ,
\ee
which arises as a component of the light-cone $\mathcal{N} = 4$ chiral superfield $\Phi_{\mathcal{N} = 4}$
\cite{BelDerKorMan04}
\be
\Phi_{\mathcal{N} = 4} (z, \theta^A)|_{\theta^2 = \theta, \theta^3 = 0}
=
{\dots} + \theta^1 \theta^4 \Phi (Z)
\, ,
\ee
depending on four superspace Grassmann coordinates $\theta^A$, $A=1,2,3,4$. Here we have made the following
identifications $X = \bar\phi_{14}$ and $\psi = \lambda^3$ with components of the $\mathcal{N} = 4$ fields.
In the multicolor limit, the sector is spanned by single-trace non-local operators in the light-cone superspace
\be
\label{SuperspaceLCoperator}
\mathbb{O} (Z_1, Z_2, \dots, Z_L)
=
{\rm tr}
\{
\Phi (Z_1) \Phi (Z_2) \Phi (Z_3) \dots \Phi (Z_L)
\}
\, .
\ee
Expanding these operators in Taylor series with respect to bosonic and fermionic coordinates we get
conventional Wilson operators of different field contents with an arbitrary number of covariant
derivatives $\mathcal{D}_+ =  \partial_+$ acting on $X$ and $\psi$,
\be
\mathcal{O}
=
{\rm tr}
\{
\partial_+^{k_1} X (0) \partial_+^{k_2} \psi (0) \partial_+^{k_3} \psi (0)
\dots
\partial_+^{k_L} X (0)
\}
\, .
\ee
Since the number of covariant derivatives is not restricted from above, the representations of $SL(2|1)$
to which these states belong are necessarily infinite-dimensional.

The $\mathcal{N} = 1$ superfield $\Phi (Z)$ transforms in the infinite-dimensional chiral representation
$\mathbb{V}_j$ of superconformal spin $j = \ell + b = 1$ of the $SL(2|1)$ algebra\footnote{Throughout this
paper, for an exception of a few places, we use notations and conventions of Ref.\ \cite{BelDerKorMan06}.}.
Therefore, the $L-$field operator $\mathbb{O} (Z_1, Z_2, \dots, Z_L)$ belongs to the tensor product
$( \mathbb{V}_j )^{\otimes L}$. The eigenstates of the $SL(2|1)$ spin chain belong to this space
and they can be classified according to irreducible $SL(2|1)$ representations entering the tensor
product. The corresponding local Wilson operators are known as superconformal operators
$\widehat{\mathcal{O}}$. However, to discuss them efficiently, it is convenient to pass from the
basis of operators to superconformal polynomials $\Psi$. This can be achieved by means of the
$SL(2|1)$ invariant scalar product, which projects out the superspace operator $\mathbb{O}$ to
superconformal primaries $\widehat{\mathcal{O}}$ \cite{BelDerKorMan06}
\be
\widehat{\mathcal{O}}_{\bit{\scriptstyle\alpha}} = \int \prod_{k = 1}^L [\mathcal{D} Z_k]_j
\,
\overline{\Psi_{\bit{\scriptstyle\alpha}} (Z_1, Z_2, \dots, Z_L)} \mathbb{O} (Z_1, Z_2, \dots, Z_L)
\, ,
\ee
where $\bit{\alpha}$ summarizes quantum numbers of the states and the $SL(2|1)-$invariant measure reads
\be
\int [\mathcal{D} Z]_j
=
\frac{\Gamma (j-1)}{\pi}
\int_{|z| \leq 1} d^2 z \int d \bar\theta \, d\theta
(1 - \bar{z} z + \bar\theta \theta)^{j - 1}
\, .
\ee
Once we find the lowest weight vectors $\Psi_{\bit{\scriptstyle\alpha}}$ in these representations, the
remaining eigenstates can be obtained from $\Psi_{\bit{\scriptstyle\alpha}}$ by applying the raising
operators $V^+$ and $\bar{V}^+$ of the superalgebra. The Bethe Ansatz and Baxter equation give the spectrum
of the lowest weights only. Being the lowest weights, the eigenstates $\Psi_{\bit{\scriptstyle\alpha}}$
diagonalize the operators $J$ and $\bar{J}$ of the Cartan subalgebra and the quadratic Casimir
operator $\mathbb{C}_2$ acting in $( \mathbb{V}_j )^{\otimes L}$
\be\label{C23}
\mathbb{C}_2 \Psi_{\bit{\scriptstyle\alpha}} = J \bar J \Psi_{\bit{\scriptstyle\alpha}}
\, , \qquad
J \Psi_{\bit{\scriptstyle\alpha}} = (m + L) \Psi_{\bit{\scriptstyle\alpha}}
\, , \qquad
\bar{J} \Psi_{\bit{\scriptstyle\alpha}} = \bar{m} \Psi_{\bit{\scriptstyle\alpha}}
\, ,
\ee
such that the lowest weights are parametrized by the vector of quantum numbers $\bit{\alpha} = [L, \bar{m}, m]$
with $m$ and $\bar m$ being nonnegative integers. These integers define the transformation properties
of the eigenstates under dilatations $L^0 = \ft12 (J + \bar{J})$ and $U(1)$ rotations $B = \ft12 (J -
\bar{J})$,
\be
\label{m-mbar}
\Psi_{\bit{\scriptstyle\alpha}} (\lambda^2 z, \lambda \theta)
=
\lambda^{m + \bar m} \Psi_{\bit{\scriptstyle\alpha}} (z, \theta)
\, , \qquad
\Psi_{\bit{\scriptstyle\alpha}} (z, \lambda^{-1} \theta)
=
\lambda^{m - \bar m} \Psi_{\bit{\scriptstyle\alpha}} (z, \theta)
\, ,
\ee
where $(z,\theta) \equiv \{z_k,\theta_k|1\le k\le L\}$ denotes the coordinates in the light-cone
superspace. In other words, $(m+\bar{m})/2$ defines the scaling dimension of the eigenstates while
$m - \bar{m}$ defines its $U(1)$ charge.

There are natural restrictions on possible values of the integers $m$ and $\bar{m}$. Examining the
transformation properties of $\Psi_{\bit{\scriptstyle\alpha}}$ under \re{m-mbar}, one finds that
$\bar m-m \ge 0$. For $m - \bar{m} =0$ the wave function $\Psi_{\bit{\scriptstyle\alpha}}$ does not
depend on $\theta$'s and it is a function of $z-$variables only. Since $\Psi_{\bit{\scriptstyle\alpha}}$
is the lowest weight, it has also to be annihilated by the lowering operators $V^-$ and $\bar{V}^-$ of
the algebra. This leads to $\Psi_{\bit{\scriptstyle\alpha}} = 1$ or, equivalently, $m = \bar{m} = 0$.
Below we will choose it as a pseudovacuum state in the nested Bethe Ansatz which does not have any Bethe
roots associated with it and possesses vanishing energy. It corresponds to the local Wilson operator (up
to an overall normalization)
\be
\widehat{\mathcal{O}}_{[L,0,0]} = {\rm tr} X^L (0)
\, .
\ee

The expansion of the single-trace operators \re{SuperspaceLCoperator} in powers of `odd' variables
truncates at order $L$ and reads
\be\label{TaylorGrassmannExp}
\mathbb{O}_L (Z) = X_L (z) + \ldots + \psi_L (z)\,\prod_{k=1}^L \theta_k\,,
\ee
where $X_L (z)=\tr \left[X (z_1) \ldots X (z_L)\right]$ and $\psi_L (z) = \tr\left[ \psi(z_1)\ldots
\psi(z_L) \right]$ are the lowest and highest components, respectively. The highest component in the
expansion \re{TaylorGrassmannExp} possesses the maximal $U(1)$ charge $L$. However, it is a
descendant of the lowest weight vector
\be\label{vec1}
\Psi_{\bit{\scriptstyle\alpha}}(z, \theta)= \theta_{12} \theta_{23} \ldots \theta_{L-1,L}
\chi^{(1)}_L (z) \sim V^- \theta_1 \ldots \theta_L \, \chi^{(1)}_L (z)
\, ,
\ee
which is proportional to a homogeneous polynomial in $\theta$'s of degree $(L-1)$ found from the requirement
that $\Psi_{\bit{\scriptstyle\alpha}}$ should be annihilated by the lowering operators $V^- = \sum_k
\partial_{\theta_k}$ being the lowering operator in $(\mathbb{V}_{j})^{\otimes L}$. Here $\chi^{(1)}_L (z)$
is a translation invariant function of $z_k$ (with $k=1,\ldots,L$) and we use the convention $\theta_{jk} =
\theta_j - \theta_k$. Thus its $U(1)$ charge is $\bar{m} - m = L - 1$. For instance, for $L = 2$, there is just
one lowest state $\Psi = \theta_{12} \chi^{(1)}_2 (z)$ with $\chi^{(1)}_2 (z) = \chi^{(1)}_2 (z_1 - z_2)$
\cite{BelDerKorMan05}. The above lowest state corresponds to the operator possesses the field content
${\rm tr} [X \psi]$. The remaining operators of the supermultiplet with different particle content are
deduced from this one by applying the step-up fermionic operators $V^+$ and $\bar{V}^+$. For $\chi
(z_1 - z_2) = z_{12}^n$, we get conventional two-particle conformal operators $\widehat{\mathcal{O}}_{
[2,n,n-1]} = {\rm tr} [X (\partial_{[+]})^n P^{(0,1)}_n (\partial_{[-]}/\partial_{[+]}) \psi]$ with
$\partial_{[\pm]} = \partial_{1,+} \pm \partial_{2,+}$.

We conclude then that the possible values of integers $m$ and $\bar m$ are subject to the constraint
$1 \le \bar{m} - m \le L - 1$. The sum $m + \bar{m}$, on the other hand, is unrestricted from above since
local Wilson operators can carry an arbitrary number of covariant derivatives.

For the low boundary $\bar{m} - m = 1$, the eigenstate $\Psi_{\bit{\scriptstyle\alpha}}$ has a unit
$U(1)$ charge and, therefore, it is given by a linear combination of $\theta$'s with prefactors depending
on $z-$variables only. The latter are fixed from the requirement that $\Psi_{\bit{\scriptstyle\alpha}}$
has to be annihilated by the lowering generators yielding
\be\label{vec2}
\Psi_{\bit{\scriptstyle\alpha}}(z, \theta)
=
\bar{V}^- \chi^{(2)}_L (z) \,,
\ee
with $\chi^{(2)}_L (z)$ being yet another translation invariant function of $z_k$ (with $k=1,\ldots,N$) and
$\bar{V}^- =\sum_k \theta_k \partial_{z_k}$ being the lowering operator in $(\mathbb{V}_{j})^{\otimes L}$.
Generally, the states with $\bar{m} - m = 1$ is related to the $L-$scalar operator with $\bar{m}$ derivatives,
while $\bar{m} - m = L - 1$ to $L-$gaugino operators with $m$ derivatives as their descendants, i.e.,
\be
\widehat{\mathcal{O}}_{[L, \bar{m}, \bar{m} - 1]} \sim \tr [ \partial_+^{\bar{m}} X^L (0) ]
\, , \qquad
\widehat{\mathcal{O}}_{[L, L + m - 1, m]} \sim \tr [ \partial_+^{m} \psi^L (0) ]
\, ,
\ee
respectively. The dilatation operator mixes together different components of the sum
carrying the same number of $\theta$'s. A distinguished feature of the two components,
$X_L(z)$ and $\psi_L(z)$, is that the dilatation operator acts on them autonomously. For
such states, corresponding to the so-called maximal helicity operators~\cite{BraDerMan98}.

The case $2 \le \bar m - m\le L-2$, is realized for the spin chain of length $L \ge 4$. The eigenstate
$\Psi_{\bit{\scriptstyle\alpha}}$ carries the $U(1)$ charge equal to $\bar m - m$ and it is given by a
homogeneous polynomial in $\theta$'s of degree $\bar m - m$ with the coefficient given by $z-$dependent
functions. In distinction to \re{vec1} and \re{vec2}, these functions are, in general, independent of
each other. For $L=4$, the lowest weight with $\bar{m} - m = 2$ is
\ba
\Psi_{[4, \bar{m}, \bar{m} - 2]}
\!\!\!&=&\!\!\!
\left(
\theta_{12} \theta_{23} \partial_{23}
+
\theta_{12} \theta_{24} \partial_{24}
+
\theta_{13} \theta_{34} \partial_{34}
\right) \chi^{(3)}_{L = 4} (z)
\, ,
\ea
where $\chi$ are translation-invariant functions of the coordinates $z$ and we have also introduced
notations for $\partial_{jk} = \partial_j - \partial_k$.

\section{Bethe Ansatz and Baxter equation}

As we pointed out in introduction, the $SL(2|1)$ integrable spin chains based on $R-$matrices can
be solved via either nested Bethe Ansatz \cite{Yan67} or Baxter approach \cite{Bax80}. The former
relies on the existence of a pseudovacuum state in the quantum space of the model. It provides a
solution to the energy spectrum of the model and leads to expressions for the eigenvalues of transfer
matrices in terms of two sets of Bethe roots. The transfer matrices is the main ingredient of the
Baxter approach, with the Baxter operators themselves being certain transfer matrices with a
special--spectral parameter-dependent---dimension of representations in the auxiliary space. However,
since one lacks a systematic procedure to construct long-range integrable spin chains corresponding
to gauge theories, one therefore has to resort to techniques which bypass the microscopic treatment
and rely on general properties of macroscopic systems. The analytical Bethe Ansatz method, which is
a generalization of the inverse transfer matrix method, was developed to determine the spectrum of
transfer matrices for closed chains \cite{Res84} and serves the purpose. In this approach, one uses
general properties of the $R$-matrix, such as analyticity, unitarity, crossing symmetry, etc., to derive
various properties of the transfer-matrix eigenvalues. These properties are used to completely
determine the eigenvalues, assuming that they have the form of "dressed" pseudovacuum eigenvalues.
We will solely concentrate on the eigenvalues of the transfer matrices, not their eigevectors.
Thus we assume this approach based on conjectured form of nested Bethe Ansatz equations at higher
orders of perturbation theory and subsequent use of the so-called analytical Bethe Ansatz to find
transfer matrices.

\subsection{Short-range magnet}

The $SL(2|1)$ spin chain has in fact three different pseudovacuum states and, as a consequence,
one can construct three different nested Bethe Ansatz solutions \cite{EssKor92,BelDerKorMan06}.
We choose however a single nesting corresponding to the first level vacuum built from scalars $X$
and as a second (nested) vacuum we use the primary excitation $\mathcal{D}_+ X$. The nested Bethe
Ansatz equations for the superconformal spin $j = 1$ of the site read
\be
\label{LONABA}
\left( \frac{u_{0,k}^+}{u_{0,k}^-} \right)^L
=
\prod_{j \neq k = 1}^{\bar{m}} \frac{u_{0,k}^- - u_{0,j}^+}{u_{0,k}^+ - u_{0,j}^-}
\prod_{l = 1}^{\bar{m} - m - 1}
\frac{u_{0,l}^+ - u_{0,j}^{(1)}}{u_{0,l}^- - u_{0,j}^{(1)}}
\, , \qquad
1
=
\prod_{j = 1}^{\bar{m}}
\frac{u_{0,k}^{(1)} - u_{0,j}^+}{u_{0,k}^{(1)} - u_{0,j}^-}
\, ,
\ee
where $u^\pm = u \pm \ft{i}{2}$. A distinguished feature of this noncompact model as compared with
conventional compact spin chains is that the total spin can now take arbitrarily large values
and the energy spectrum of the model is not restricted from above even for a finite length of the
spin chain. For specific low values of $[L, \bar{m}, m]$ the set of transcendental equations \re{LONABA}
can be solved numerically. These studies show that all first and second level Bethe roots are real.
The one-loop anomalous dimensions are determined by the first level Bethe roots $u_{0,k}$ only via
the formula
\be
\gamma^{(0)}
= \frac{i}{2}
\sum_{j = 1}^{\bar{m}} \left( \frac{1}{u_{0,j}^+} - \frac{1}{u_{0,j}^-} \right)
\, .
\ee

Let us introduce two polynomials parametrized by the first and second level Bethe roots
\be
Q_0 (u) = \prod_{k = 1}^{\bar{m}} (u - u_{0,k})
\, , \qquad
Q_0^{(1)} (u) = \prod_{k = 1}^{\bar{m} - m - 1} \left( u - u^{(1)}_{0,k} \right)
\, .
\ee
These are eigenvalues of the Baxter operators of the $SL(2|1)$ spin chain. As is well known the number
of independent Baxter functions depends on the rank of the symmetry group. In the present case there
are three. However, only two of them are polynomial in the spectral parameter $u$, with the remaining
one being a meromorphic function of $u$. All transfer matrices of the chain can be expressed in terms
of these polynomials \cite{BelDerKorMan06}.

The transfer matrices can be uniquely fixed using the analytical Bethe Ansatz \cite{Res84} requiring
that they should be polynomial in $u$, free from poles at Bethe roots of the first and second level.
We will presently consider only matrices corresponding to lowest dimension representation in the
auxiliary space. This consideration immediately yields\footnote{The results of Ref.\ \cite{BelDerKorMan06}
are reproduced under the following redefinition of the spectral parameter, $u \to i u$, and the chiral
Baxter functions, $Q_3 (u) \to Q_0 (u - \ft{i}{2})$ and $Q^{(0)}_{13} (u) \to Q^{(1)}_0 (u)$.}
\be
\label{tau0}
\tau_0 (x)
=
(u^-)^L \frac{Q_0 (u - i)}{Q_0 (u)}
+
(u^+)^L
\left[ \frac{Q_0 (u + i)}{Q_0 (u)} - 1 \right] \frac{Q^{(1)}_0 (u^-)}{Q^{(1)}_0 (u^+)}
\, .
\ee
It corresponds to the eigenvalues of the transfer matrices with atypical three dimensional
representations $(1)_+$ in the auxiliary space. One can also construct the transfer matrix
with atypical representation $(1)_-$ in the auxiliary space and it reads
\be
\bar{\tau}_0 (x)
=
(u^+)^L \frac{Q_0 (u + i)}{Q_0 (u)}
+
(u^-)^L
\left[ \frac{Q_0 (u - i)}{Q_0 (u)} - 1 \right] \frac{Q^{(1)}_0 (u^+)}{Q^{(1)}_0 (u^-)}
\, .
\ee
These transfer matrices are polynomials of order $L$ in the spectral parameter $u$ with coefficients
determined by the conserved charges of the chain,
\be
\label{LOtransferMatrices}
\tau_0 (u) = (u^-)^L + \sum_{k = 2}^L q_{0,k} (u^-)^{L - k}
\, , \qquad
\bar{\tau}_0 (u) = (u^+)^L + \sum_{k = 2}^L \bar{q}_{0,k} (u^+)^{L - k}
\, ,
\ee
and the leading charge determined by the eigenvalues of the quadratic superconformal Casimir
operator $q_{0,2} = \bar{q}_{0,2} = \mathbb{C}_2 = \bar{m} (m + L)$. Eliminating the auxiliary
Baxter operator $Q^{(1)}_0$ from the transfer matrices, one finds a second order finite-difference
equation for the Baxter function $Q_0$ \cite{BelKor05,BelDerKorMan06},
\be
\left[  \tau_0 (u) \bar{\tau}_0 (u) - (u^+ u^-)^L \right] Q_0 (u)
=
(u^+)^L \left[ \tau_0 (u) - (u^-)^L \right] Q_0 (u + i)
+
(u^-)^L \left[ \bar{\tau}_0 (u) - (u^+)^L \right] Q_0 (u - i)
.
\ee
By solving this polynomial equation in the spectral parameter $u$, one finds the roots of the Baxter
function in terms of the conserved charges $q_{0,k}$ and $\bar{q}_{0,k}$ as well as the quantized values
of the latter. Once found, the Baxter function determines the eigenspectrum of the anomalous dimensions
at one loop,
\be
\label{GammaLO}
\gamma^{(0)} = \ft{i}{2} \left( \ln Q_0 (\ft{i}{2}) - \ln Q_0 (- \ft{i}{2}) \right)^\prime
\, .
\ee

A distinguished feature of the lowest $X_L(z)$ and highest $\psi_L(z)$ components of the superspace operator
\re{SuperspaceLCoperator} is that the $SL(2|1)$ dilatation operator acts on them autonomously. As we
previously established, the state $X_L (z)$ is a descendant of the $SL(2|1)$ lowest weight vector \re{vec2}
with $\bar m - m = 1$ while $\theta_1 \ldots \theta_L \psi_L (z)$ is a descendant of the lowest weight
\re{vec1} with $\bar m - m = L - 1$. In both cases, the $SL(2|1)$ Hamiltonian effectively reduces to the
Hamiltonian of the $SL(2)$ spin chain of length $L$ and spins $\ell = \ft12$ and $\ell = 1$, respectively.

\subsection{Long-range magnet}

We now turn to multi-loop generalization of the $SL(2|1)$ Baxter equation. The starting point of our
consideration is a generalization of Bethe Ansatz equations for the sector to all orders in 't Hooft
coupling constant $g = g_{\rm\scriptscriptstyle YM} \sqrt{N_c}/(2 \pi)$. Since the Bethe equations were
conjectured for the full gauge theory in \cite{BeiSta05}, we can get the ones for the minimally
supersymmetric subsector by removing excitations not belonging to the sector in question. One immediately
finds that the $SL(2|1)$ nested Bethe Ansatz equations take the following form
\ba
\label{AllONABA}
\left( \frac{x_k^+}{x_k^-} \right)^L
\!\!\!&=&\!\!\!
\prod_{j \neq k = 1}^{\bar{m}} \frac{x_k^- - x_j^+}{x_k^+ - x_j^-}
\frac{
\left( 1 - \frac{g^2}{4 x_k^+ x_j^-} \right)
}{
\left( 1 - \frac{g^2}{4 x_k^- x_j^+} \right)
}
{\rm e}^{i \theta (x^+_k, x_j) - i \theta (x^-_k, x_j)}
\prod_{l = 1}^{\bar{m} - m - 1}
\frac{x_l^+ - x_j^{(1)}}{x_l^- - x_j^{(1)}}
\, , \\
1
\!\!\!&=&\!\!\!
\prod_{j = 1}^{\bar{m}}
\frac{x_k^{(1)} - x_j^+}{x_k^{(1)} - x_j^-}
\, .
\ea
They are written in terms of the renormalized spectral parameter $x = \ft12 (u + \sqrt{u^2 - g^2})$
\cite{BeiDipSta04} using the convention $x^\pm = x [u^\pm]$ and embody a nontrivial magnon-magnon
scattering phase factor\footnote{We slightly changed notations used in Ref.\ \cite{BeiEdeSta06} to
accommodate $\theta$ for our needs.} $\theta$ which is indispensable to have agreement with string
theoretical calculations \cite{AruFroSta04,BeiHerLop06} and four-loop calculations of cusp anomalous
dimension in maximally supersymmetric gauge theory \cite{BerCzaDixKosSmi06} as was recently demonstrated
in Ref.\ \cite{BeiEdeSta06},
\be
\theta (x_k^\pm, x_j)
=
4 \sum_{m = 1}^\infty \sum_{n = 0}^\infty
(- 1)^{m + n} \mathcal{Z}_{2m, 2n+1} (g)
\left[
\frac{\mathcal{Q}_{2n + 1} (x_j)}{(x^\pm_k)^{2 m}} - \frac{\mathcal{Q}_{2 m} (x_j)}{(x^\pm_k)^{2 n + 1}}
\right]
\, .
\ee
It is expressed in terms of the single-excitation charges
$\mathcal{Q}_k (x)$,
\be
\mathcal{Q}_k (x_j) = \frac{1}{(x^+_j)^k} - \frac{1}{(x^-_j)^k}
\, ,
\ee
and expansion coefficients depending on the coupling constant
\be
\mathcal{Z}_{m, n} (g)
= \left( \frac{g}{2} \right)^{m + n} \int_0^\infty d t
\frac{J_m (g t) J_n (g t)}{t ({\rm e}^{t} - 1)}
\, .
\ee
The anomalous dimensions are determined by the first level Bethe roots $x_k$ via
\be
\gamma (g)
= \frac{i g^2}{2}
\sum_{j = 1}^{\bar{m}} \left( \frac{1}{x_j^+} - \frac{1}{x_j^-} \right)
\, .
\ee
In the zero 't Hooft coupling limit, the long-range equations \re{AllONABA} and the anomalous dimensions
$\gamma (g)/g^2$ naturally reduce to one-loop Bethe Ansatz \re{LONABA} and \re{GammaLO} of the previous
section. Analyses based on the $SL(2)$ reduction, i.e., $\bar{m} - m = 1$, of the long-range equations
demonstrated \cite{KotLip06,BenBenvKleSca06} correct interpolation of cusp anomaly to strong coupling
predictions from string theory for the energy of dual rotating string configuration on the anti-de Sitter
space \cite{GubKlePol02,FroTse02,Kru02,BelGorKor06,SakSat06}.

\begin{figure}[t]
\begin{center}
\mbox{
\begin{picture}(0,180)(90,0)
\put(0,0){\insertfig{6}{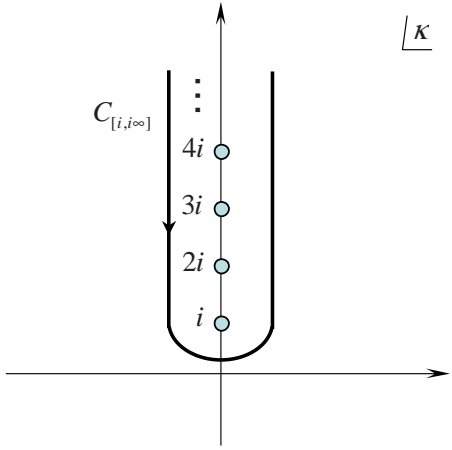}}
\end{picture}
}
\end{center}
\caption{\label{KappaContour} The integration contour in the magnon scattering
phase.}
\end{figure}

Now, analogously to the one-loop discussion of the previous section, we introduce two
Baxter polynomials build from the first and second level Bethe roots
\be
Q (u) = \prod_{k = 1}^{\bar{m}} (u - u_k (g))
\, , \qquad
Q^{(1)} (u) = \prod_{k = 1}^{\bar{m} - m - 1} \left( u - u^{(1)}_k (g) \right)
\, ,
\ee
which admit perturbative expansion to all orders in coupling constant $u_k (g) = u_{0,k}
+ g^2 u_{1,k} + \dots$ and analogously for $u^{(1)}_k (g)$. The transfer matrix can be
constructed using the analytical Bethe Ansatz \cite{Res84} requiring pole free structure
at Bethe roots of the first and second level. One gets
\ba
\label{tau}
\tau (x)
\!\!\!&=&\!\!\!
(x^-)^L {\rm e}^{\Delta_- (x^-)} \frac{Q(u - i)}{Q (u)}
\\
&+&\!\!\!
(x^+)^L
\left[
{\rm e}^{\Delta_+ (x^+)}
\frac{Q(u + i)}{Q (u)}
-
{\rm e}^{\ft12 \Delta_- (x^+) + \ft12 \Delta_+ (x^+)}
\right]
\frac{Q^{(1)} (u^-)}{Q^{(1)} (u^+)}
{\rm e}^{\ft12 \sigma_0^{(1)} (x^-) - \ft12 \sigma_0^{(1)} (x^+)}
\, , \nonumber
\ea
which is a multi-loop generalization of the transfer matrices with $(1)_+$ representation in
the auxiliary space. Here we introduced the convention
\be
\Delta_\pm (x) = \sigma_\pm (x) - \Theta (x)
\, ,
\ee
for the difference of the trivial dressing factor, represented in terms of the Baxter
polynomial \cite{Bel06},
\be
\prod_{j = 1}^{\bar{m}} \left( 1 - \frac{g^2}{4 x x_j^\pm} \right)
=
{\rm e}^{- \frac{1}{2} \sigma_\mp (x)}
\, .
\ee
with
\be
\label{AllLoopSigma}
\sigma_\eta (x)
=
\int_{- 1}^1 \frac{dt}{\pi} \,
\frac{\ln Q \left( \eta \ft{i}{2} - g t \right)}{\sqrt{1 - t^2}}
\left(
1 - \frac{\sqrt{u^2 - g^2}}{u + g t}
\right)
\, ,
\ee
and the magnon scattering phase factor
\be
\prod_{j = 1}^{\bar{m}} {\rm e}^{i \theta (x, x_j)}
=
{\rm e}^{\Theta (x)}
\, ,
\ee
which takes a more involved form
\ba
\label{Theta}
\Theta (x)
\!\!\!&=&\!\!\!
g \int_{- 1}^1 \frac{d t}{\sqrt{1 - t^2}}
\ln \frac{Q \left( - \ft{i}{2} - g t \right)}{Q \left( \ft{i}{2} - g t \right)}
\,
{-\!\!\!\!\!\!\int}_{-1}^1 ds \frac{\sqrt{1 - s^2}}{s - t}
\nonumber\\
&\times&
\int_{C_{[i, i \infty]}} \frac{d \kappa}{2 \pi i} \frac{1}{\sinh^2 (\pi \kappa)} \
\ln
\left( 1 + \frac{g^2}{4 x x [\kappa + g s]} \right)
\left( 1 - \frac{g^2}{4 x x [\kappa - g s]} \right)
\, .
\ea
The integration contour in the variable $\kappa$ is represented in Fig.\ \ref{KappaContour}. Equation
\re{tau} reduces to the one-loop transfer matrix \re{tau0} of the previous section for $g = 0$. One
can also construct the transfer matrix with antichiral representation in the auxiliary space,
\ba
\label{taubar}
\bar{\tau} (x)
\!\!\!&=&\!\!\!
(x^+)^L {\rm e}^{\Delta_+ (x^+)} \frac{Q(u + i)}{Q (u)}
\\
&+&\!\!\!
(x^-)^L
\left[
{\rm e}^{\Delta_- (x^-)}
\frac{Q(u - i)}{Q (u)}
-
{\rm e}^{\ft12 \Delta_+ (x^-) + \ft12 \Delta_- (x^-)}
\right]
\frac{Q^{(1)} (u^+)}{Q^{(1)} (u^-)}
{\rm e}^{\ft12 \sigma_0^{(1)} (x^+) - \ft12 \sigma_0^{(1)} (x^-)}
\, . \nonumber
\ea
Similarly to Eqs.\ \re{LOtransferMatrices}, the transfer matrices \re{tau} and \re{taubar} are given by
the power series but in the renormalized spectral parameter with the expansion coefficients determined
by the integrals of motion which, in turn, admit and infinite series representation in 't Hooft coupling,
\be
\tau (x)
=
(x^-)^L \left( 1 + \sum_{k \geq 1} \frac{q_k (g)}{(x^-)^k} \right)
\, , \qquad
\bar{\tau} (x)
=
(x^+)^L \left( 1 + \sum_{k \geq 1} \frac{\bar{q}_k (g)}{(x^+)^k} \right)
\, .
\ee
In contrast to one loop, the all-order transfer matrices acquire nontrivial coefficients
in front of $x^{L - 1}$. These can be found explicitly by studying the large-$u$ limit of
\re{tau} and \re{taubar}. The first subleading asymptotics of the transfer matrices
immediately yields $q_1$ and $\bar{q}_1$,
\be
q_1 (g) = \bar{q}_1 (g)
=
- \frac{g^2}{2} \int_{-1}^1 \frac{d t}{\pi} \sqrt{1 - t^2}
\left(
\ln Q \left( \ft{i}{2} - g t \right) + \ln Q \left( - \ft{i}{2} - g t \right)
\right)^\prime
-
\theta_1 (g)
\, ,
\ee
where $\theta_1 (g)$ is the coefficient of the leading $1/u-$asymptotics of the magnon phase
\re{Theta}.

In order to obtain the Baxter equation for the polynomial $Q(u)$, one eliminates the auxiliary
Baxter operator $Q^{(1)}$ from the transfer matrices, and finds a second order finite difference
equation, which we call by analogy with the one-loop case, the long-range Baxter equation
\ba
\label{SL21LongRangeBaxter}
&&
\left[
\tau(x) \bar{\tau} (x)
-
(x^+ x^-)^L
{\rm e}^{
\ft12 \Delta_+ (x^-) + \ft12 \Delta_+ (x^-)
+
\ft12 \Delta_- (x^+) + \ft12 \Delta_- (x^-)
}
\right] Q (u)
\\
&&\qquad\qquad\qquad
=
(x^+)^L {\rm e}^{\Delta_+ (x^+)}
\left[
\tau (x)
-
(x^-)^L
{\rm e}^{\ft12 \Delta_+ (x^-) + \ft12 \Delta_- (x^-)}
\right] Q (u + i)
\nonumber\\
&&\qquad\qquad\qquad
+ \,
(x^-)^L {\rm e}^{\Delta_- (x^-)}
\left[
\bar{\tau} (x)
-
(x^+)^L
{\rm e}^{\ft12 \Delta_+ (x^+) + \ft12 \Delta_- (x^+)}
\right] Q (u - i)
\nonumber \, .
\ea
The anomalous dimensions to all orders in gauge coupling constant are determined by the equation
\be
\label{AllOrderAD}
\gamma (g) = i g^2 \int_{- 1}^1 \frac{dt}{\pi} \sqrt{1 - t^2}
\left(
\ln Q \left( \ft{i}{2} - g t \right)
-
\ln Q \left( - \ft{i}{2} - g t \right)
\right)^\prime
\, .
\ee

Working our perturbative expansion in coupling constant of both sides of Eq.\ \re{SL21LongRangeBaxter}
and anomalous dimension \re{AllOrderAD},
\be
\gamma (g) = \sum_{k \geq 0} g^{2 (k + 1)} \gamma^{(k)}
\, ,
\ee
one can explicitly solve for eigenvalues with small $m$ and $\bar m$. For the low boundary $\bar{m}
- m = 1$, we recover the bosonic $SL(2)$ long-range Baxter equation \cite{BelKorMul06,Bel06} albeit with
a nontrivial magnon scattering phase and reproduce all available multi-loop gauge theory calculations
\cite{KotLipVel04,EdeJarSok04,Ede04,EdeSta06,BelKorMul06,BerCzaDixKosSmi06} and predictions based on
integrability \cite{BeiDipSta04,Sta04,Zwi05}. For the upper boundary $\bar{m} - m = L - 1$, the solution
to the Baxter equation for $L = 3$ site chains agree with field-theoretical calculations of the
three-gaugino anomalous dimensions performed in Ref.\ \cite{BelKorMul05} and for other values of $L$
with the algebraically constructed two-loop dilatation operator in Ref.\ \cite{Zwi05}. A few specific
eigenvalues for up to four loops are displayed in Table \ref{ExactSpectra} with the magnon phase stepping
in at order $\mathcal{O} (g^8)$.

\begin{table}
\renewcommand{\arraystretch}{1.5}
\begin{center}
\begin{tabular}[pos]{||c|c|c|c|c||}
\hline \hline
$\bit{\alpha}$ & $\gamma^{(0)}$ & $\gamma^{(1)}$ & $\gamma^{(2)}$ & $\gamma^{(3)}$
\\
\hline \hline
$[3, 5, 3]$
&
$5$
&
$- \ft{245}{48}$
&
$\ft{21475}{2304}
$
&
$-$
\\
\hline
$[3, 7, 5]$
&
$\ft{133}{24}$
&
$- \ft{131117}{23040}$
&
$\ft{1039405829}{99532800}$
&
$-$
\\
\hline
$[4, 5, 2]$
&
$\ft{32 \pm \sqrt{10}}{6}$
&
$- \ft{9334 \pm 269 \sqrt{10}}{1728}$
&
$\ft{48971080 \pm 1339361 \sqrt{10}}{4976640}$
&
$- \ft{5 (1259661488 \pm 33839563 \sqrt{10})}{286654464}
-
\ft{905 \pm 23\ \sqrt{10}}{640} \zeta (3)$
\\
\hline $[4, 6, 3]$
&
$\ft{17}{3}$
&
$- \ft{5005}{864}$
&
$\ft{658771}{62208}$
&
$-\ft{423834365}{17915904} - \ft{443}{288} \zeta(3)$
\\
\hline \hline
\end{tabular}
\end{center}
\caption{\label{ExactSpectra} Eigenvalues of selected states up to four-loop order in $\mathcal{N}=4$
super-Yang-Mills theory.}
\end{table}

The Baxter equation can be immediately used to find the Sudakov behavior of the anomalous dimensions
for asymptotically large values of the quadratic Casimir $\mathbb{C}_2$. As was established in Ref.\
\cite{BelGorKor06,EdeSta06}, the minimal anomalous dimension is independent of the twist of Wilson
operators since while $q_{0,2} = \bar{q}_{0,2} = \mathbb{C}_2$ takes large values along the trajectory,
all other integrals of motion become anomalously small such that the spectral curve determining their
asymptotics degenerates into one of twist-two operators. An analysis along the lines of Ref.\
\cite{Bel06} yields equations for the cusp anomalous dimension of the lowest trajectory in the
spectrum in the large$-\mathbb{C}_2$ limit,
\be
Q (u) Q_0 (u \pm i) = {\rm e}^{\Delta_\pm (x^\pm)} Q (u \pm i) Q_0 (u)
\, .
\ee
This results immediately implies universality of the cusp anomalous dimension for operators with different
field content in agreement with independence of the soft gluon radiation of the spin of elementary fields
it is emitted from. The fine structure of the spectrum determined by subleading expansion coefficients
depends on the particle content of composite operators and deserves a dedicated study.

\section{Outlook}

Presently we have applied the method of the Baxter $\mathbb{Q}-$operator which plays the central r\^ole
in the method of separated variables to minimally supersymmetric subsector of the maximally supersymmmetric
gauge theory. Within the formalism, the eigenvalue $Q_0 (u)$ of $\mathbb{Q}$ is identified with a
single-particle wave function which obeys a Schr\"odinger equation which coincides with the Baxter
equation. We have constructed a generalization of the Baxter equation for the graded $SL(2|1)$ magnet to
all-orders in 't Hooft coupling. This putative spin chain arises in the closed $SL(2|1)$ subsector of the
$\mathcal{N} = 4$ supersymmertric Yang-Mills theory. The TQ-relation was found to admit a second order
finite-difference form with coefficients determined by ``dressed'' fundamental atypical transfer matrices.
The main advantage of the above construction is that it can be straightforwardly generalized to other subsectors
of the $\mathcal{N} = 4$ super-Yang-Mills theory based on noncompact supergroups of higher rank. This
immediately applies to single-trace operators with suppressed particle-number changing transitions, with
the $SL(2|2)$ subsector of the $\mathcal{N} = 4$ SYM being its maximal sector. However, above it the length
changing effects take place and one has to modify the formalism to accommodate them within the Baxter approach.
Analytical Bethe Ansatz can be naturally used to construct transfer matrices with higher dimensional
representations in the auxiliary space. Let us point out that the bilinear combination of transfer matrices
in the left hand side of the Baxter equation is related to the transfer matrix $t_{n/2, b}$ with typical
$4n-$dimensional representation $(b, n/2)$ in the auxiliary space for $n = 2, b = 0$ \cite{BelDerKorMan06}.
One can construct the fusion hierarchy of transfer matrices to determine higher-dimensional transfer matrices
along the lines of Ref.\ \cite{Tsu88}. A step in this direction has been recently undertaken in Ref.\ \cite{Bei06}.

Still, to identify the underlying long-range spin chain, one needs the explicit form of the $\mathbb{Q}-$operator.
Acting on the Wilson operators in the superspace representation \re{SuperspaceLCoperator}, the one-loop Baxter
operator can be realized as an integral operator acting on the positions of superfields in superspace, in a close
analogy with the dilatation operator which arises as a coefficient in the expansion of its kernel in the spectral
parameter \cite{BelDerKorMan06}. The open question remains to find operator representation for higher order Baxter
functions and ultimately to all orders in coupling.

\vspace{0.5cm}

This work was supported by the U.S. National Science Foundation under grant no. PHY-0456520.



\end{document}